\documentclass[prl,showpacs,twocolumn]{revtex4}
\usepackage{graphicx}
\usepackage{psfig}

\begin{document}

\newcommand {\ee}[1] {\label{#1} \end{equation}}
\newcommand{\be}{\begin{equation}}
\newcommand {\e} {\varepsilon}
\def\reff#1{(\ref{#1})}
\def\S{\mbox{\bf S}}
\def\P{\mbox{\bf P}}
\def\I{\mbox{\bf I}}
\def\re{\mbox{Re}}
% \draft
% \tightenlines

% \wideabs{
\title{Noise-induced dynamics in bistable systems with delay}
\date{\today}
\author{L. S. Tsimring}
\affiliation{Institute for Nonlinear Science, University of California, 
San Diego, La Jolla, CA 92093-0402}
\author{A. Pikovsky}
\affiliation{Department of Physics, University of Potsdam,
	Postfach 601553, D-14415 Potsdam, Germany}

\begin{abstract}%
Noise-induced dynamics of a prototypical bistable system with delayed
feedback is studied theoretically and numerically. For small noise and
magnitude of the feedback, the problem is reduced to the analysis of the
two-state model with transition rates depending on the
earlier state of the system. In this two-state approximation, we found
analytical formulae for the autocorrelation function, the power
spectrum, and the linear response to a periodic perturbation.  They show
very good agreement with direct numerical simulations of the original
Langevin equation. The power spectrum has a pronounced peak at the
frequency corresponding to the inverse delay time, whose amplitude has a
maximum at a certain noise level, thus demonstrating coherence
resonance.  The linear response to the external periodic force also has
maxima at the frequencies corresponding to the inverse delay time and
its harmonics.  
\end{abstract}

% }
%\twocolumn
\maketitle
The effects of random noise on bistable systems and the related
phenomenon of stochastic resonance have received an enormous attention
in the last decade. As a result, a comprehensive theory and a whole
range of experimental observations have emerged (for a recent review 
see~\cite{sr}).  In many physical as well as biological systems, the
time-delayed feedback plays a significant role in the dynamics.  These
systems in the absence of noise have been well thoroughly investigated
using the theory of delay-differential equations (DDE)~\cite{dde}. The
theory of stochastic delay-differential equations (SDDE), in which
effects of noise and time delay are combined, remains much less studied. 
Meanwhile, it appears that the combination of these features is
ubiquitous in nature as well as in society. Examples include biophysiological
dynamics~\cite{longtin90-chen97} and laser dynamics in optical cavities 
\cite{garcia96-buldu01-masoller01}. It is also believed that the
combined effects of noise, bistability, and delay play an important role
in gene regulatory networks \cite{collins00}.

The delayed stochastic systems have been a subject of several recent
papers~\cite{longtin,ohira95-00,ohira99}.  In
Refs.~\cite{longtin}, a systematic statistical description
of a certain class of stochastic delay-differential equations was
developed in the limit of small time delay. More interesting, however,
is the case of a large time delay which is comparable with the mean
Kramers transition time determined by the noise intensity and the
potential barrier height. In this case, resonant phenomena may occur
which  would lead to spontaneous oscillations of the system with a certain
preferred frequency. In Refs.~\cite{ohira95-00}, Ohira and 
co-workers studied the related phenomenon of delayed random walks. 
In that model, the hopping
probability depends on the position of the particle a given number of
hops in the past. In certain cases, the particle diffusion is limited,
and it exhibits quasi-regular oscillations near the origin. 
In Ref.~\cite{ohira99},  Ohira and Sato studied a discrete-time 
two-state system in which the occupancy probabilities of
the two states depended on the state of the system some $N$ time steps
before. While that model also showed some interesting resonant features, it
appears to us somewhat unrealistic, since the states of the system at
two consecutive iterations are completely uncoupled, and its dynamics is
in fact identical to that of a superposition of  $N$ independent
one-dimensional maps affected by random noise.  In most practically
relevant cases, however, the state of the system should be affected in
the first place by its immediate past, with additional correction
arising from the time-delayed feedback.

In this Letter we study the effects of the thermal activation on
bistable systems with additional time-delayed feedback. 
Our prototypical model is the overdamped particle motion in the double-well
quartic potential $U(x(t),x(t-T))$, described by the Langevin equation
\begin{eqnarray}
\frac{dx(t)}{dt}&=&
-\frac{\partial U(x(t),x(t-T))}{\partial x(t)}+\sqrt{2D}\xi(t)
\nonumber\\
&\equiv&
x(t)-x^3(t)+\epsilon x(t-T)+\sqrt{2D}\xi(t).
\label{sdde}
\end{eqnarray}
Here $T$ is the delay and $\epsilon$ is the strength of the feedback, and
 $\xi(t)$ is a Gaussian white noise with $\langle \xi\rangle=0$ 
and $\langle \xi(t)\xi(t')\rangle=\delta(t-t')$.

In our analytic description we approximate (\ref{sdde}) with a two-state
(dichotomic) system, in
which the dynamical variable $s(t)$ takes two values $s=\pm 1$. This
reduction has
been successfully used in studies of the stochastic resonance \cite{mcnamara89}. The
dynamics of $s$ is fully determined by the switching rates, i.e. by the
probabilities to switch $s\rightarrow -s$. Because of the delay, we have two
switching rates depending on the state $s(t-T)$: $p_1$ if the state at time $t-T$ is
the same as at time $t$, and $p_2$ otherwise. Thus, the the switching rate can
be written as
\begin{equation}
p(t)=\frac{p_1+p_2}{2}+\frac{p_1-p_2}{2} s(t)s(t-T)\;.
\label{rates}
\end{equation}
A quantitative relation between the rate process (\ref{rates}) and the original 
model (\ref{sdde}) can be easily established for small $D$ and $\epsilon$ by
virtue of the Kramers formula for the escape rate \cite{kramers}
$r_K=(2\pi)^{-1}\sqrt{U''(x_m)U''(x_0)}\exp[-\Delta U/D]$, where $x_m$
and $x_0$ are the positions of the minimum and the maximum of the potential,
respectively, and $\Delta U$ is the potential barrier to cross over. 
For small $D$, the switching rates are small compared to the intra-well
equilibration rate, and the probability 
density distribution is close to a narrow Gaussian distribution centered around
$x_m$, and so the adiabatic approximation applies. 
For small $\epsilon$, $|x_m|=1\pm \epsilon/2$ depending on the sign of
$x(t)x(t-T)$, $x_0=0$, and in the first order in $\epsilon$ we obtain
\begin{equation}
p_{1,2}=
\frac{\sqrt{2\pm 3\epsilon}}{2\pi}\exp\left[-\frac{1\pm 4\epsilon}{4D}\right]\;.
\label{trans}
\end{equation}

Without loss of generality let us assume that the system is at state
$s=1$ at time $0$. 
We define $n_\pm(t)$ to be the probability of attaining value $\pm 1$ 
at time $t$. The master equations for $n_\pm(t)$ is written in a usual
way,
\begin{equation}
\begin{array}{l}
\dot{n}_+(t)=-W_\downarrow (t) n_+(t) + W_\uparrow (t) n_-(t)\;,\\[1ex]
\dot{n}_-(t)=-W_\uparrow (t) n_-(t) + W_\downarrow (t) n_+(t)\;,
\end{array}
\label{master1}
\end{equation}
where $W_\uparrow (t)\;dt$ is the probability of transition from $-1$ to
$+1$ within time interval $(t,t+dt)$
and vice versa. In our stochastic model with time-delayed feedback,
\begin{equation}
\begin{array}{l}
W_\downarrow(t)= p_1 n_+(t-T) + p_2 n_-(t-T)\;,\\[1ex]
W_\uparrow(t)= p_2 n_+(t-T) + p_1 n_-(t-T)\;.
\end{array}
\label{wpm}
\end{equation} 
Substituting (\ref{wpm}) in (\ref{master1}) and making use
of the normalization condition $n_-(t) + n_+(t) = 1$, we obtain
\begin{eqnarray}
\dot{n}_+(t)= p_2 n_-(t) - p_1 n_+(t) - (p_2-p_1)n_-(t-T),
\label{master2}\\
\dot{n}_-(t)= - p_1 n_-(t) + p_2 n_+(t) - (p_2-p_1)n_+(t-T).
\label{master3}
\end{eqnarray} 
The correlation function $C(\tau)$ is determined as 
%the averaged time
%response function
\begin{equation}
C(\tau)=\langle s(\tau)s(0)\rangle=\langle s(\tau)\rangle=n_+(\tau)-n_-(\tau),
\end{equation}
(we recall that the initial state is $s(0)=1$).
Thus, replacing $t$ with $\tau$ and subtracting (\ref{master3}) from 
(\ref{master2}), we obtain
\begin{equation}
\frac{dC(\tau)}{d\tau}=-(p_1+p_2)C(\tau)+(p_2-p_1)C(\tau-T).
\label{ceq2}
\end{equation}
This equation should be complemented with the symmetry $C(-\tau)=C(\tau)$ and the
normalization $C(0)=1$ conditions.

The solution of this linear equation on the interval $(0,T)$ 
can be found using 
ansatz $C(\tau)=A\exp(-\lambda \tau) + B \exp\lambda(\tau-T)$. 
Plugging this ansatz in Eq.(\ref{ceq2}) yields
$\lambda=2\sqrt{p_1p_2}, \
B=A(\sqrt{p_2}-\sqrt{p_1})/(\sqrt{p_2}+\sqrt{p_1})\exp(-2\sqrt{p_1p_2}T)$. 
Constant $A$ is found from the condition $C(0)=1$, and we
obtain
\begin{equation}
C(\tau)=\frac{(\sqrt{p_1}+\sqrt{p_2})e^{-\lambda\tau}+
(\sqrt{p_2}-\sqrt{p_1})e^{\lambda(\tau-T)}}
{\sqrt{p_1}+\sqrt{p_2}+(\sqrt{p_2}-\sqrt{p_1})e^{-\lambda T}}\;.
\label{ctauf}
\end{equation}
Using (\ref{ceq2}) and (\ref{ctauf}), one can easily calculate 
$C(\tau)$ at all $\tau>T$,
\begin{eqnarray}
&&C(nT+\tau')=e^{-(p_1+p_2)\tau'}C(nT)+(p_2-p_1)\times\nonumber\\
&&\times\int_0^{\tau'}
C((n-1)T+t)e^{(p_1+p_2)(t-\tau')}dt\;,
\label{ctauf1}
\end{eqnarray}
where $n=1,2,...$ and $0<\tau'<T$.

\begin{figure}[!htb]
  \centerline{\includegraphics[width=0.43\textwidth]%
  {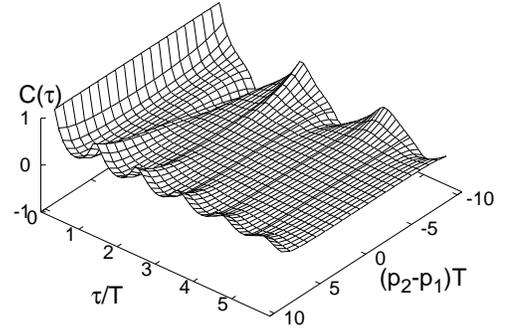}}
  \caption{The autocorrelation function in the two-state model as a
function of 
  the time lag $\tau$ and the feedback strength
  ${p}_2-{p}_1$, for $({p}_1+{p}_2)T=10$.}
  \label{fig:corf}
\end{figure}
We present correlation function (\ref{ctauf}),(\ref{ctauf1}) as a function
of  normalized time $\tau/T$ and dimensionless parameters 
$p_{1,2}T$ in Fig.~\ref{fig:corf}. Its structure
differs depending on whether the feedback is positive ($p_2>p_1$, what
corresponds to a positive $\epsilon$ in (\ref{sdde})), or negative ($p_2<p_1$,
$\epsilon<0$). For positive feedback the correlation function is positive, and has
maxima at $\tau\approx nT$.  For negative feedback the peaks at $\tau\approx nT$ have alternating
signs. It is interesting to note that the peaks of the correlation
function are always delayed with respect to $nT$. For $\lambda
T\gg 1$, the time interval corresponding to the first peak is
\begin{equation}  
\tau_1=T+(\sqrt{p_2}-\sqrt{p_1})^{-2}\ln\frac{(\sqrt{p_1}+\sqrt{p_2})^2}{2(p_1+p_2)}\;.
\label{offset}
\end{equation}

 From the correlation function we can also determine the power spectrum
$S(\omega)=\int_{-\infty}^\infty C(\tau)\cos(\omega\tau)\,d\tau$.
 It is
convenient to derive the expression for $S$ directly from equation (\ref{ceq2}).
Denoting $L(\omega)=\int_0^\infty C(\tau)\exp[i\omega\tau]\,d\tau$ and
substituting here (\ref{ceq2}) we obtain
\begin{equation}
-C(0)-i\omega L=-(p_1+p_2)L+(p_2-p_1)e^{i\omega T}
\left[L-I(\omega)\right]\;,
\label{psp1}
\end{equation}
where 
\begin{widetext}
\[ 
I(\omega)=\int_0^T C(\tau)e^{-i\omega\tau} d\tau=\frac{(\sqrt{p_1}+\sqrt{p_2})[1-e^{(-i\omega -\lambda)T}]
(i\omega+\lambda)^{-1}+
(\sqrt{p_2}-\sqrt{p_1})
[e^{-i\omega T}-e^{-\lambda T}](i\omega -\lambda)^{-1}}
{\sqrt{p_1}+\sqrt{p_2}+(\sqrt{p_2}-\sqrt{p_1})e^{-\lambda T}}\;.
%\label{psp2}
\]
\end{widetext}
Using $C(0)=1$ and $S(\omega)=2\re L(\omega)$, we obtain 
%from (\ref{psp1}) finally
\begin{equation}
S(\omega)=2\re\frac{1+(p_2-p_1)e^{i\omega
T}I(\omega)}{p_1+p_2-(p_2-p_1)e^{i\omega T}-i\omega}\;.
\label{pow1}
\end{equation}

We compare this analytic result with numerical simulations of the bistable
oscillator (\ref{sdde}) in Fig.~\ref{fig:spec}.

\begin{figure}[!htb]
  \centerline{\includegraphics[width=0.38\textwidth]%
  {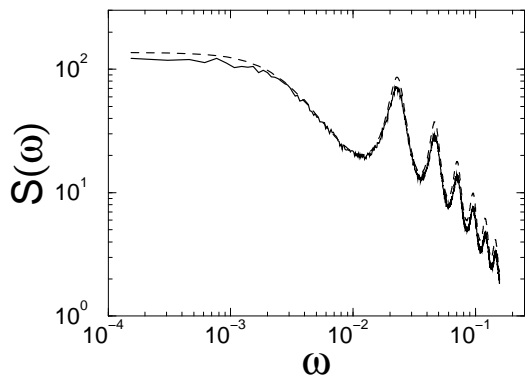}}
  \caption{Comparison of the power spectrum in Eq.~(\ref{sdde}) for $D=0.1$,
  $T=250$, $\epsilon=0.05$ (solid line) with theory~(\ref{pow1}) (dashed line). 
  }
  \label{fig:spec}
\end{figure}
\begin{figure}[!htb]
  \centerline{\includegraphics[width=0.42\textwidth]%
  {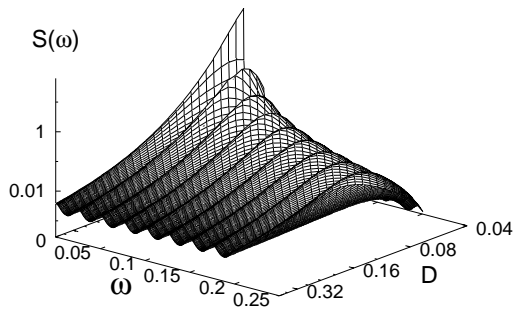}}
%  {spectr_bw.eps}}
  \caption{Power spectrum in the delay system (\ref{sdde}) calculated in the
  two-state approximation for $T=250$ and $\epsilon=0.1$. 
  }
  \label{fig:spec2}
\end{figure}
In figure~\ref{fig:spec2} we show the dependence of the power spectrum on the
noise intensity $D$, while the feedback parameter $\epsilon$ is kept constant
(the switching rates $p_{1,2}$ are calculated according to (\ref{trans})). One
can see that the peak at the main frequency $\omega\approx 2\pi/T$ has a maximum
at a certain noise level. This is a characteristic feature of the coherence
resonance \cite{pik97}: the coherence in the noise-driven system attains maximum
at a ``resonant'' noise temperature. In the present case the underlying 
physical mechanism is the resonance between the Kramers rate and the delay. 
If the Kramers rate is small (for
small noise intensity), a characteristic interval between the switches is larger
than delay time, and the latter is not displayed in the spectrum because the
process is a purely Poissonian one 
(with a renormalized due to the feedback switching
rate). For an intermediate Kramers rate the switchings are highly influenced by
the feedback, with a preferable periodicity with the delay time $T$ being
manifested as a peak in the spectrum. For large noise intensity the effect of
the feedback decreases again, because the relative magnitude of the delayed 
feedback $(p_2-p_1)/(p_1+p_2)$ is proportional to $D^{-1}$.

Let us discuss now the response of the time-delay stochastic system to a periodic
external force. Similarly to Ref.\cite{mcnamara89},
we assume that the transition rates (\ref{wpm}) are modulated
with a frequency $\Omega$ according to the Arrhenius rate law,
\begin{eqnarray*}
W_\downarrow (t)&=& [p_1 n_+(t-T) + p_2 n_-(t-T)]e^{\gamma(t)}
\;,\\
W_\uparrow (t)&=& [p_2 n_+(t-T) + p_1 n_-(t-T)]e^{-\gamma(t)}
\;,
\end{eqnarray*} 
where $\gamma(t)=\mu D^{-1}\cos(\Omega t +\phi)$.
The equation for the quantity $\sigma(t)=n_+(t)-n_-(t)$ (which now is not the
autocorrelation function) now reads
\begin{eqnarray*}
\frac{d\sigma}{dt}&=&-(p_1+p_2)(n_+e^{\gamma(t)}-n_-e^{-\gamma(t)})\nonumber
\\&&+
(p_2-p_1)(n_+e^{\gamma(t)}+n_-e^{-\gamma(t)})\sigma(t-T)\;.
%\label{clr}
\end{eqnarray*}
In the linear approximation $\mu<<1$ this reduces to
\[
\frac{d\sigma}{dt}=-(p_1+p_2)(\sigma+\gamma(t))+
(p_2-p_1)\sigma(t-T)(1+C(t)\gamma(t))\;.
%\label{clr1}
\]
Now writing $\sigma=\sigma_0+\mu D^{-1} \sigma_1$ where $\sigma_0$ is the solution
(\ref{ctauf},\ref{ctauf1}), we obtain for the first-order correction $\sigma_1$
\begin{eqnarray}
&&\frac{d\sigma_1}{dt}=-(p_1+p_2)\sigma_1+(p_2-p_1)\sigma_1(t-T)\nonumber\\&&
+
((p_2-p_1)\sigma_0(t)\sigma_0(t-T)-(p_1+p_2))\cos(\Omega t +\phi)\;.
\label{clr2}
\end{eqnarray}
We are interested in the response at the frequency $\Omega$ for $t\to\infty$,
because only this part contributes to the delta-peak in the spectrum at this
frequency. For $t\to\infty$ $\sigma_0(t)\to 0$, so we can neglect the corresponding
term $(p_2-p_1)\sigma_0(t)\sigma_0(t-T)$ in (\ref{clr2}) and write the solution as
\[
\sigma_1(t)=
\mbox{Re}\frac{(p_1+p_2)e^{i\Omega t +\phi}}{(p_2-p_1)e^{-i\Omega T}-i\Omega
-(p_1+p_2)}\;.
%\label{clr3}
\]
This is exactly the periodic component at frequency $\Omega$ 
in the process $s(t)$, and the linear
response $\eta$ is
\begin{equation}
\eta=\frac{1}{2D^2}\frac{(p_1+p_2)^2}{\left|(p_2-p_1)e^{-i\Omega T}-i\Omega
-(p_1+p_2)\right|^2}\;.
\label{clr5}
\end{equation}
In the  absence of delayed feedback, when $p_1=p_2=r_K$, this expression coincides with
that of \cite{mcnamara89} for the stochastic resonance in the two-state
model. With feedback, the response demonstrates a resonance-like structure
in dependence on the driving frequency (contrary to the classical stochastic
resonance), see Fig.~\ref{fig:resp}.
\begin{figure}[!htb]
  \centerline{\includegraphics[width=0.38\textwidth]%
  {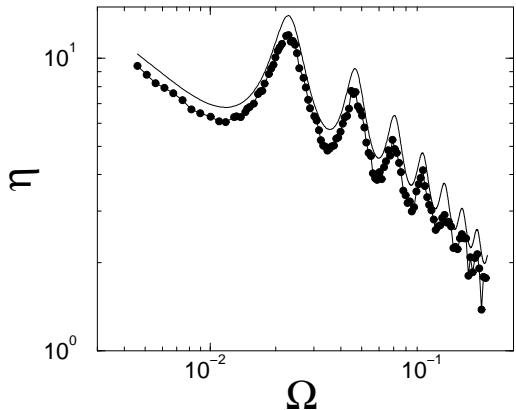}}
  \caption{Linear response of model (\ref{sdde}) for $D=0.1$,
  $T=250$, $\epsilon=0.05$, normalized by the variance of the process 
  (circles), compared with  theory~(\ref{clr5}) (line)
  }
  \label{fig:resp}
\end{figure}

In conclusion, we have developed a theory of a noise-driven bistable
system with delayed feedback. In general, this problem is very
difficult because of the non-Markovian nature of the dynamics. However,
for small noise and small magnitude of the feedback, the problem can be
greatly simplified by reduction to the two-state (dichotomic) model with
certain transition rates which depend on the earlier state of the
system. In this two-state approximation, we were able to derive the
analytical formulae for the autocorrelation function, the power
spectrum, and the linear response to a periodic perturbation.  They show
very good agreement with direct numerical simulations of the
corresponding Langevin equation. The power spectrum has a pronounced
peak at the frequency corresponding to the delay time, whose amplitude
has a maximum at a certain noise level, thus demonstrating coherence
resonance. This level corresponds to the mean switching time comparable
to the delay time.  The linear response to the external periodic force
also has maxima at the frequencies corresponding to the inverse delay
time and its harmonics.

In a more general context of multistable dynamical systems with memory,
the behavior of the system depends on its past through some memory
kernel. Such a kernel is equivalent to multiple time delays. A similar
analysis of the correlation properties for such systems would be of
great interest.  Furthermore, in applications, multiple feedback loops
with different delay times occur in networks of interacting elements,
such as biological neurons, stock traders, or Internet nodes.  It is
very interesting to study the influence of noise on the dynamics of
such networks. We anticipate the emergence of spontaneous oscillations and
the coherence resonance features similar to the effects considered in
this Letter.

We thank A. Neiman and M. Rosenblum for helpful discussions.
L.T. acknowledges support from the U.S. Department of Energy, 
Engineering Research Program of the Office of Basic Energy Sciences
under grants DE-FG03-95ER14516 and DE-FG03-96ER14592, and from the
U.S. Army Research Office under MURI grant DAAG55-98-1-0269.

\end{document}